# Multiprocessor Scheduling Using Parallel Genetic Algorithm

Nourah Al-Angari[1], Abdullatif ALAbdullatif[2]

[1,2] Computer Science Department, College of Computer & Information Sciences, King Saud University
Riyadh, Saudi Arabia

**Abstract**
Tasks scheduling is the most challenging problem in the parallel computing. Hence, the inappropriate scheduling will reduce or even abort the utilization of the true potential of the parallelization. Genetic algorithm (GA) has been successfully applied to solve the scheduling problem. The fitness evaluation is the most time consuming GA operation for the CPU time, which affect the GA performance. The proposed synchronous master-slave algorithm outperforms the sequential algorithm in case of complex and high number of generations' problem.

## 1. Introduction

Parallel computing is a promising approach to meet the increased need for high speed machines. Furthermore, tasks scheduling is the most challenging problem regarding parallel computing. Hence, the inappropriate scheduling will reduce or even abort the utilization of the true potential of the parallelization. The scheduler goal is to assign tasks to available processors by considering the precedence constraints between tasks whilst minimizing the overall execution time "make span". This optimization problem classified as NP-complete problem. This problem becomes a strong NP-hard problem in the case of arbitrary number of processors and arbitrary task processing time.
Genetic algorithm (GA) which is a meta-heuristic algorithm has been successfully applied to solve the scheduling problem. The fitness evaluation is the most time consuming GA operation for the CPU time, which affects the GA performance. This paper proposes and implements a synchronous master-slave parallelization where the fitness evaluated in parallel.
The rest of paper organized as follow: genetic algorithm, parallel genetic algorithm, proposed algorithm, theoretical analysis, practical analysis, and conclusion.

## 2. Genetic Algorithm

Genetic algorithm introduced by John H. Holland 1975 [1] is a stochastic optimization algorithm that is basically builds on principles of natural selection and recombination. GA initialize the initial population randomly and calculate the fitness for each individual then select parent, do crossover mutation operation, calculate fitness for each new individual, finally select the next generation and repeat this process until a termination criteria or up to predefined number of generations.

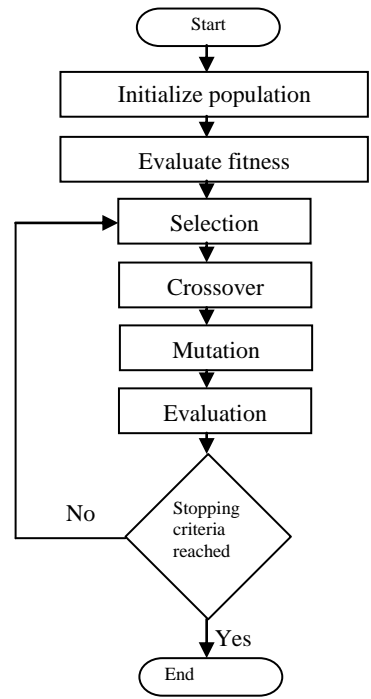

Fig. 1: GA Flow Diagram

## 3. Parallel Genetic Algorithm

There are two main possible methods to parallelism. The first of which is data parallelism, where the same instruction will be executed on numerous data simultaneously. The second one is control parallelism which involves execution of various instructions concurrently [2].

Data parallelism is sequential by its nature as only the data manipulation is paralyzed whilst the algorithm will be executed as the sequential one instruction in certain time. Thus, the majority of parallel genetic algorithms were data based parallelism.

GAs can be parallelized depending on the following:
- How to evaluate the fitness and how to apply the genetic operators
- Is it a single population or multiple subpopulations?
- In case of multiple subpopulations how the individuals will be exchanged
- How to apply the selection (local/global)

Depending on how to answer the above questions, we can categorize the PGA into:

3.1 Global single-population master-slave GAs parallelization GPGA (also known as distributed fitness evaluation)

In this algorithm there is a single population (as in sequential GA) but the fitness evaluation and/or application of genetic operators are distributed among different processors. It works as follows, the master stores the population and assigns a fraction of the population to each slave processor where they evaluate and return the fitness values and may apply some of the genetic operators fig. (2).

GPGA can be synchronous if the master waits for all slaves to finish their tasks before proceeding to the next generation. The synchronous GPGA has the same properties as sequential GA and explore the search space precisely as the sequential but it is faster.
The synchronous is easy to implement as there is no need to change the serial GA structure and it shall give a significant speedup. However, since all processors need to wait for the slowest processor that may lead to a bottle-neck that has been overcomes by the asynchronous GPGA.

In an asynchronous GPGA the master never stops to wait for the slow processors. Instead, it will use some selection operation (as tournament selection) to cover the missing individuals. Thus it becomes different than the serial GA [3] [4] [5].

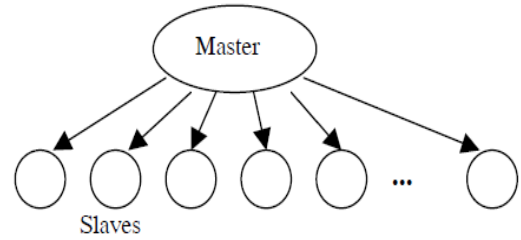

Fig. 2: master-slaves GA

3.2 Single-population fine-grained

It is suitable for the massively parallel computers so it sometimes called massively parallel genetic algorithm. It has one population that is spatially-structured usually in 2 dimensional grids which restricts the selection and mating in a small neighborhood. It introduce a new parameters the size and shape of neighborhoods see fig. (3) [3] [4] [5].

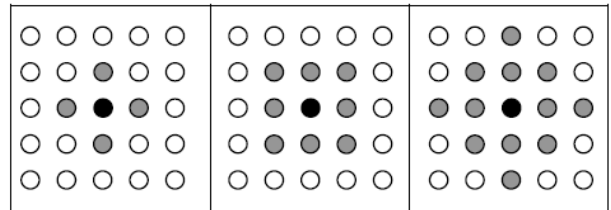

Fig. 3: single population fine-grained different parameters

3.3 Multiple-deme parallel GA

It is the most popular parallel method and it also known as a multiple-population coarse-grained GA, distributed GA, and island model. It has several subpopulations (demes). These demes are exchanging the individuals occasionally, commonly known as the migration. The size and the number of demes, the connection topology between demes fig. (4), migration rate, migration frequency, and the policy to select emigrants and to replace the existing individuals with the coming migrants are all new parameters introduced by this algorithm and affect the quality and efficiency of the algorithm [3] [4].

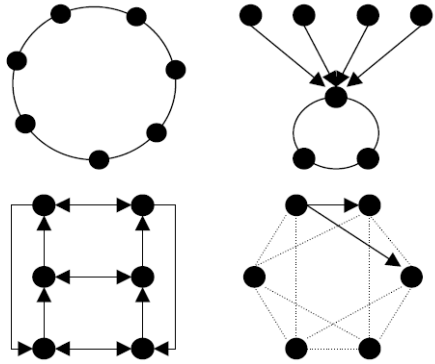

Fig. 4: Multiple-deme parallel GA

In addition to these three main models, there are other models that are combination of the above three models or combine the PGA with other optimization methods:

*Hierarchical Parallel genetic Algorithms*
HPGA combines two models of PGA to parallelize the GA. This combination forms a hierarchy [4].

*Hybrid parallel genetic algorithms*
It combines the PGA with some optimization methods e.g. hill-climbing [4].

## 4. The Proposed Approach

The proposed parallel genetic algorithm is based on the sequential algorithm on [6] and uses synchronous master-slave GAs parallelization. The master-slave GAs parallelization never affects the behavior of the algorithm, so there is no need for huge modifications and it has the full advantage of searching the whole of search space.

A master is the main processor store the full population of chromosomes and assigns a certain fraction of the individuals to slave processors, where the slaves evaluate fitness value for the assigned fraction and return their values. The parallelized done only on the fitness evaluation as the fitness of an individual is independent from the rest of the population, and there is no need to communicate during this phase.

### 4.1 Chart

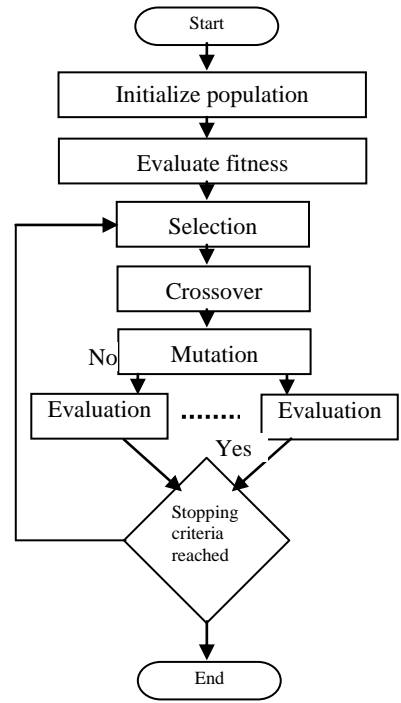

Fig. 5: Proposed approach Flow Diagram

### 4.2 Code

```
Master thread
{
initialize population;
while(termination criterion is not reached)
  {
      With probability Pc
      Parents selection
      child=crossover(survived
      individuals);
      replace  deleted  individual  with
      child;
      perform  mutation  with  probability
      pM;
      for(i=1;i<NUMBER_OF_PROCESSORS;i++)
      { create new Slave thread; }
  }
}

Slave thread{evaluate individual;}
```

## 5. Theoretical Analysis

Genetic algorithm is a stochastic algorithm where its complexity is depending on genetic operators' implementation, chromosome representation, population size, number of generation, and clearly the fitness function. Changing the previous factors has a very significant effect on the overall complexity.

The general sequential time complexity of genetic algorithm is

$$O(p * G * O(Fitness) * ((P_c * O(crossover)) + (P_m * O(mutation)))) \quad (1)$$

Where:
  *P: Population size*
  *G: number of Generations*
  *PC: Crossover probability*
  *Pm: Mutation probability*

In case of synchronous master-slave GA, P, G, $P_c$, $P_m$, O(mutation), O(crossover), and O(fitness) are constant so for simplification we will only consider the fitness in addition to the communication cost.

Sequential time complexity:

$$O(P * G * O(fitness)) \quad (2)$$

Parallel time complexity:

$$O\left(P * G * \frac{O(fitness)}{nP}\right) + CC \quad (3)$$

Where

  *nP: number of processors.*
  *CC: communication cost*

Speedup:

$$[O(P * G * O(fitness))] / [O(P * G * \frac{O(fitness)}{nP})]$$
$$= nP - CC \quad (4)$$

Cost:

$$nP * (O\left(P * G * \frac{O(fitness)}{nP}\right) + CC)$$
$$= O(P * G * O(fitness)) + CC * nP$$
$$= O(P * G * O(fitness)) \quad (5)$$

The cost is equal to the sequential time so the proposed parallel algorithm is cost optimal.

## 6. Practical Analysis

This algorithm has been implemented using java threads in a shared memory environment.

For analysing the performance of the algorithm, different strategies have been used:

1- Changing the population size
Number of iterations (generations) = 2

Table 1: the effect of the population size

| Population size | Seq. GA | Parallel GA |
|---|---|---|
| 10000 | 301 | 302 |
| 500000 | 6470 | 6904 |
| 600000 | 8842 | 8929 |

*the results represent the average run among 10 runs. And the time is in sec.

In small problem size the sequential algorithm outperforms the parallel algorithm.

2- Increasing the number of iterations (generations)
Number of processors = 2, and number of tasks =8 (simple)

Table 2: the effect of population size and number of iterations (generations)

| No. of iteration | Pop size | Seq. GA | Parallel GA |
|---|---|---|---|
| 1000 | 1000 | 32 | 32 |
| 10000 | 10000 | 1910 | 1890 |

*the results represent the average run among 10 runs. And the time is in sec.

When the problem size increased the parallel outperform the sequential time.

3- Increasing the problem complexity
Population size =10000, Number of processors=15, and number of tasks =18.

Table 3: the effect of increase the size and complexity of the problem

| No. of iteration | Seq. GA | Parallel GA |
|---|---|---|
| 1000 | 257 | 229 |
| 5000 | 466 | 381 |
| 10000 | 3051 | 2381 |

*the results represent the average run among 10 runs. And the time is in sec.

As seen in table 3 a considerable speedup has been achieved when increasing the problem size and complexity.

## 7. Conclusions

According to the obtained results, the proposed parallel algorithm outperforms the sequential algorithm in case of complex and high number of generation problems. In smaller problems, it is not preferred to use the parallel algorithms.
Using the asynchronies may give a better performance since it will never have to wait for all processors to finish their tasks.